\documentclass{nature}
\usepackage{siunitx}
\usepackage{amsmath,amssymb}
\usepackage{bm}
\usepackage{color}
\usepackage{graphicx}
\usepackage{float}
\usepackage{verbatim}
\usepackage[english]{babel}
\usepackage{marvosym}
\usepackage[colorlinks=true,bookmarks=false,citecolor=blue,urlcolor=blue]{hyperref}
\usepackage{epstopdf}
\usepackage{caption}

\makeatletter
\let\saved@includegraphics\includegraphics
\AtBeginDocument{\let\includegraphics\saved@includegraphics}

\makeatother

\title{\centering Modulating nonlinear optical responses in 3R-MoS$_2$ Fabry--P\'erot microcavities\par}

\author{Renkang Song$^{1,5}$, Ziye Chen$^{1,5}$, Junbo Xu$^{1,5}${\Large \Letter}, Zerui Wang$^1$, Zitao Wu$^1$, Shenao Zhao$^1$, Wenhao Su$^1$, Ziheng Pan$^1$, Junho Choi$^2$, Vasily Kravtsov$^3$, Di Huang$^{1}${\Large \Letter}, Zhanshan Wang$^{1,4}$ \& Tao Jiang$^{1,4}${\Large \Letter}}

\begin{document}

\maketitle

\begin{affiliations}
\upshape
\item MOE Key Laboratory of Advanced Micro-Structured Materials, Shanghai Frontiers Science Center of Digital Optics, Institute of Precision Optical Engineering, and School of Physics Science and Engineering, Tongji University, Shanghai, China.

\item Department of Physics, Kyung Hee University, Seoul, Republic of Korea.

\item School of Physics and Engineering, ITMO University, Saint Petersburg, Russia.

\item Shanghai Institute of Intelligent Science and Technology, Tongji University, Shanghai, China.

\item These authors contributed equally: Renkang Song, Ziye Chen, Junbo Xu.

{\Large \Letter}e-mail: junboxu@tongji.edu.cn; idgnauh@tongji.edu.cn; tjiang@tongji.edu.cn
\end{affiliations}

\newpage
\begin{abstract}
\section*{Abstract}
Rhombohedrally stacked transition metal dichalcogenides such as 3R-MoS$_2$ offer an exceptional platform for nonlinear optics, naturally forming Fabry-P\'erot (FP) microcavities due to their giant dielectric contrast with the surrounding media.
However, rigorously tracking the evolution of multiple harmonic fields within these unpatterned monolithic crystals remains a fundamental challenge.
Here, we establish a self-consistent framework, spanning from linear broadband reflectance to second- and third-harmonic generation (SHG and THG), to systematically decode these nonlinear behaviors.
Moving beyond conventional models, we demonstrate that the nonlinear emission is dictated by a delicate interplay among the intrinsic material absorption, the FP effects at the fundamental frequency, as well as those at the harmonic frequencies.
When harmonic photons lie below the bandgap, weak absorption allows the nonlinear spectra to exhibit a complex modulation driven by the synergistic contribution of FP effects from both fundamental and harmonic waves. 
In stark contrast, severe intrinsic absorption of higher-energy photons heavily damps the FP effects of the harmonic fields, reducing the nonlinear response to an absorption-limited regime modulated almost exclusively by the FP effects at the fundamental frequency.
By successfully decoupling these geometric and material contributions across different harmonic orders, our findings provide a precise design paradigm for engineering next-generation van der Waals photonic architectures.

\end{abstract}

\newpage
\section*{Introduction}
Transition metal dichalcogenides (TMDCs) have emerged as compelling platforms for nonlinear optics due to their exceptionally high nonlinear susceptibilities \cite{mak2016photonics, Autere2018NLO2Dreview, zhao2025applications,hu2019coherent, jiang2014valley, trolle2015observation, klimmer2021all, zhang2024advanced, zeng2024nonlinear, chen2017enhanced}.
Unlike the widely studied 2H phase, the rhombohedral 3R phase (e.g., 3R-MoS$_2$) retains broken inversion symmetry regardless of the number of layers, enabling persistent and scalable harmonic generation well beyond the monolayer limit \cite{shi20173r, zograf2024combining, xu2022towards, song2026polarizationengineeringsecondharmonicgeneration,qin2024interfacial, xu2025spatiotemporal}.
To date, efforts to push the boundaries of nonlinear conversion efficiency in these materials have heavily relied on complex artificial engineering.
These advanced approaches include  quasi-phase matching in periodically poled structures, twist-angle phase matching via vertical assembly, and excitation of localized cavity modes or resonant modes in patterned metasurfaces \cite{trovatello2025quasi, tang2024quasi, hong2023twist, tognazzi2025interface, munson2026cavity, seidt2025ultrafast, zograf2025ultrathin, peng20253r}.

While these advanced structuring techniques are powerful, they often demand intricate fabrication processes and can obscure the fundamental optical behaviors inherent to the pristine material.
Crucially, 3R-MoS$_2$ possesses a distinct but frequently underestimated characteristic: its exceptionally high refractive index ($n \sim 4$ in the near-infrared regime)  \cite{MoS2_permittivity, mooshammer2024enabling, weber2023intrinsic, Luo2024Strong, ermolaev2021giant}.
In the thick-layer regime, this massive dielectric contrast with the surrounding environment naturally transforms an unpatterned, monolithic flake into a highly efficient Fabry--P\'erot (FP) microcavity \cite{wan2026quasi,abtahi2025thickness}.
These intrinsic FP cavity effects can significantly confine and reshape the local optical fields, creating resonance conditions that profoundly impact both linear and nonlinear optical processes.
Therefore, rigorously decoupling these geometric cavity enhancements from the intrinsic material properties is essential for accurately evaluating and fully exploiting the nonlinear potential of van der Waals microcavities.

In this work, we systematically investigate the distinct cavity-modulated behaviors of second- and third-harmonic generation (SHG and THG) in 3R-MoS$_2$ microcavities. 
We first establish a built-in optical calibration protocol via broadband reflectance spectra \cite{choi2010measurement, gillen2005use}.
This approach rigorously resolves the wavelength-scale FP effects while avoiding the morphological uncertainties often encountered with standard atomic force microscopy (AFM) for thick layers \cite{lee2002nonlinear, garcia1999attractive, tang2011hysteresis}. 
Building upon this precise in situ metrology, we show that the nonlinear signal is governed by three distinct factors: the fundamental wave FP resonance, the generated harmonic FP modes, and the intrinsic material absorption. 
Under weak material absorption (for sub-bandgap harmonic signals), the nonlinear response exhibits complex modulation driven by the interplay of both fundamental and harmonic FP modes.
In stark contrast, when the generated photon energy lies well above the material's bandgap (including both THG and short-wavelength SHG), strong material absorption heavily suppresses the harmonic cavity modes. 
Consequently, the fundamental FP modes almost dominate these nonlinear processes.
By successfully decoupling the interaction between fundamental and harmonic FP modes from intrinsic material properties, this study provides a comprehensive framework for understanding how wavelength-scale geometry shapes nonlinear optical responses, guiding the design of future van der Waals photonic architectures.

\section*{Results}

\subsection{Intrinsic Fabry--P\'erot Microcavities in monolithic 3R-MoS$_2$.}
\begin{figure*}[htbp]
    \centering
    \includegraphics[width=\textwidth]{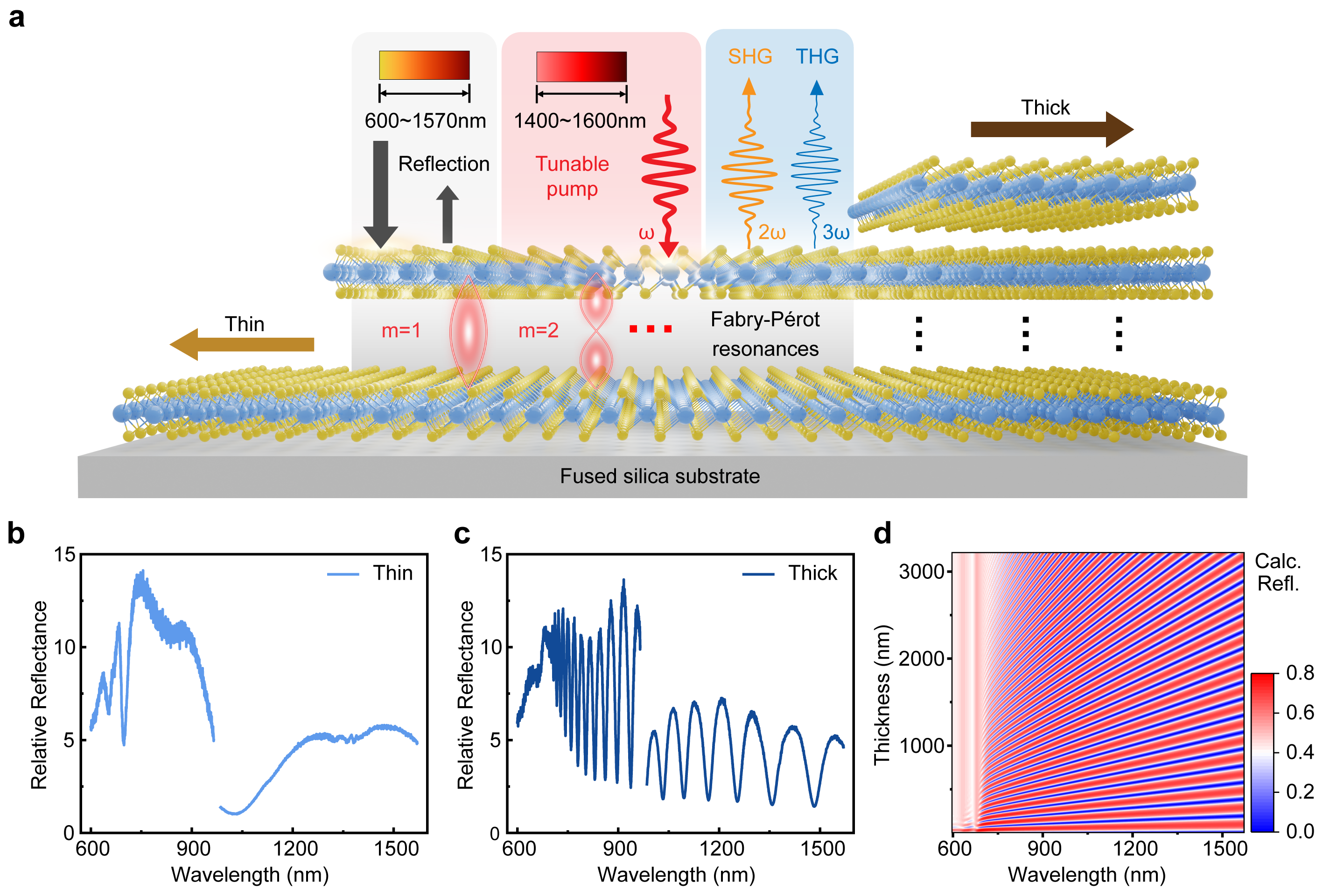} 
    \caption{
    \textbf{Concept and linear optical responses of 3R-MoS$_2$ microcavities.} 
    \textbf{a} Schematic of the physical mechanisms in a 3R-MoS$_2$ microcavity. The strong refractive index contrast among air, the 3R-MoS$_2$ flake, and the silica substrate naturally forms a FP cavity. Multiple interfacial reflections induce strong interference, fundamentally modulating both the linear reflectance and the higher-order harmonic generation (SHG, $2\omega$; THG, $3\omega$) driven by a fundamental pump field ($\omega$). 
    \textbf{b, c} Linear relative reflectance spectra of a representative thin ($\sim$ 100~nm) flake (\textbf{b}) and a thick ($\sim$ 2000~nm) flake (\textbf{c}). The pronounced periodic oscillations in the thick flake are a direct manifestation of the linear FP interference effect. 
    \textbf{d} 2D contour map depicting the global evolution of the linear reflectance spectra as a function of incident wavelength spanning from 600 to 1570~nm and flake thickness up to $\sim$3210~nm.}
    \label{fig:fig1}
\end{figure*}

The experiments were conducted on 3R-MoS$_2$ flakes supported by a fused silica (SiO$_2$) substrate.
Driven by the high refractive index contrast at the upper and lower boundaries, the flake itself naturally acts as a highly efficient FP microcavity, as conceptually illustrated in Fig.~\ref{fig:fig1}a.
These intrinsic longitudinal modes serve as a fundamental resonant framework.
They strongly reshape the optical fields, ultimately modulating the nonlinear responses of the 3R-MoS$_2$ flakes.

As a first step, we utilized a broadband halogen lamp to investigate the linear optical responses of the samples across the visible--near-infrared spectral range. 
The spectral modulation exhibits a pronounced thickness-dependent evolution: a relatively thin ($\sim$100~nm) flake (Fig.~\ref{fig:fig1}b) shows broad spectral features, whereas a much thicker flake (Fig.~\ref{fig:fig1}c) displays dense, periodic oscillations.
These distinct spectral features are direct manifestations of the FP interference effect in the linear optical regime, dictated by both the geometric cavity thickness and the material's intrinsic dispersion properties.

We established a theoretical reflectance model based on the transfer matrix method (TMM) to quantitatively resolve this mode evolution as a function of both thickness and wavelength.
Crucially, this model incorporates the material's intrinsic dispersion.
The resulting 2D contour map (Fig.~\ref{fig:fig1}d) directly reveals two distinct optical signatures.
First, we observe numerous dispersive reflection fringes that shift continuously with wavelength and thickness, a direct manifestation of FP cavity resonances.
Second, strongly enhanced reflection peaks emerge across all thicknesses at energies corresponding to the A and B excitonic resonances (typically around 681~nm and 631~nm) of 3R-MoS$_2$ \cite{Ullah2021bl3RMoS2}. 
Unlike the dispersive FP fringes, these excitonic features are intrinsically related to the material's band structure and remain independent of the geometric cavity thickness. 
By accurately capturing this interplay between cavity resonances and intrinsic material dispersion, our TMM framework provides a robust predictive foundation for accurate, non-destructive thickness calibration, as discussed in the following section.

\subsection{Precision Metrology via Optical Calibration.}

\begin{figure*}[htbp]
    \centering
    \includegraphics[width=\textwidth]{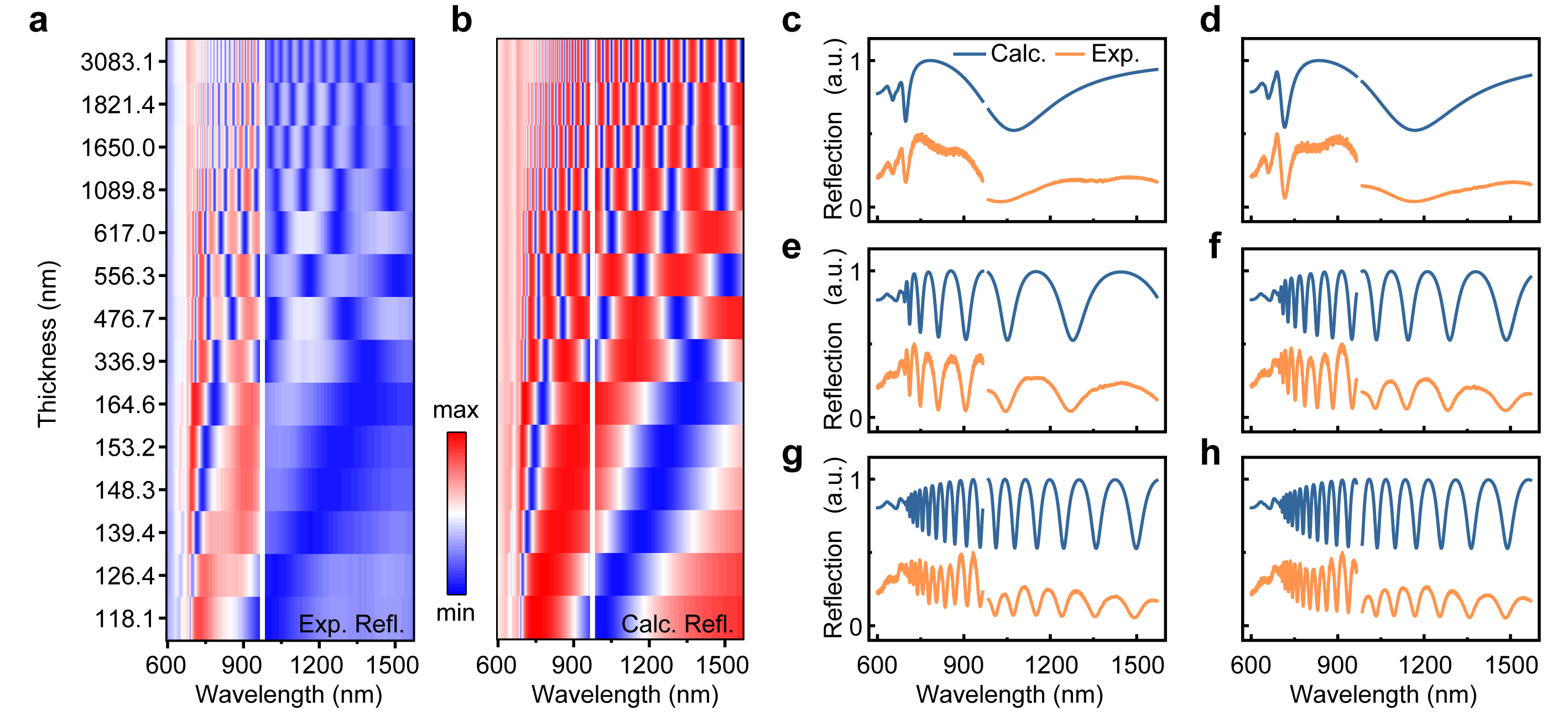}
    \caption{
    \textbf{Broadband reflectance spectra and thickness calibration for 3R-MoS$_2$.} 
    \textbf{a} Experimental two-dimensional mapping of the broadband reflectance spectra (measured continuously from 600 to 1570~nm under halogen lamp illumination) as a function of wavelength and sample thickness. The trajectory of the interference fringes clearly exhibits the FP cavity mode evolution. 
    \textbf{b} Corresponding theoretically calculated two-dimensional reflection intensity distribution based on the TMM, accurately capturing the fringe compression and mode dispersion bending behavior. 
    \textbf{c--h} Detailed spectrum-by-spectrum comparisons between the experimental measurements (orange curves) and theoretical calculations (solid blue lines) for six representative thicknesses ranging from the hundred-nanometer to the micrometer scale (126.4~nm (\textbf{c}), 344.2~nm (\textbf{d}), 617.0~nm (\textbf{e}), 1089.8~nm (\textbf{f}), 1422.3~nm (\textbf{g}), and 1821.4~nm (\textbf{h})). The data gap between 965~nm and 985~nm arises from the low-sensitivity crossover region between the visible and infrared detectors.}
    \label{fig:fig2}
\end{figure*}

To extract the accurate cavity dimensions and overcome the morphological uncertainties of standard atomic force microscopy (AFM) in the thick-layer regime (see Supplementary Information Sec.~1 for a detailed technical discussion), we established an in situ optical calibration protocol.
We systematically measured the reflectance spectra of the samples over a broad wavelength range (600 to 1570~nm).
By fitting these spectra, we extracted the precise thickness for each flake.
We then mapped the experimental reflectance spectra as a joint function of wavelength and these calibrated thicknesses, constructing a two-dimensional contour plot (Fig.~\ref{fig:fig2}a).
Within this map, the continuous spectral shift and scaling of the resonance modes with increasing thickness serve as hallmarks of robust multi-beam interference within the 3R-MoS$_2$ microcavity.

The robust global agreement of this fitting procedure is confirmed in Fig.~\ref{fig:fig2}b, where the theoretical model faithfully reproduces the experimental reflection pattern.
Specifically, it precisely captures both the progressive reduction of the free spectral range (FSR) at larger thicknesses and the dispersion-induced curvature of the resonance fringes.
To verify the model's quantitative accuracy, we extracted representative spectra for flakes ranging from the hundred-nanometer to the micrometer scale. 
This enabled a fine, spectrum-by-spectrum comparison (Fig.~\ref{fig:fig2}c--h, see Supplementary Information Sec.~3 for more results).
The experimental data points exhibit excellent agreement with the theoretical curves across all investigated thicknesses.
The model successfully reproduces the exact resonance peak positions, evolving oscillation periods, and overall spectral lineshapes.

Because the FP resonance is highly sensitive to the optical path length, fitting multiple longitudinal modes ensures that the effective geometric thickness is determined with minimal error.
This robust metrology guarantees the self-consistency of our theoretical framework across the entire thickness range.
Building on this geometric baseline, we now turn to the cavity's nonlinear responses.
Crucially, the linear FP resonances confirmed here dictate a constructive interference, leading to a profound resonant enhancement of the localized fundamental field.
Because nonlinear optical emission scales with the power of the driving field intensity, this localized enhancement translates into a magnified modulation of the final harmonic yield.
Relying on these accurate thicknesses, we next systematically investigate how these intrinsic FP resonances govern the giant modulation and complex spectral evolution of SHG.

\subsection{Complex Spectral Reshaping of Second-Harmonic Generation.}

\begin{figure*}[htbp]
    \centering
    \includegraphics[width=\textwidth]{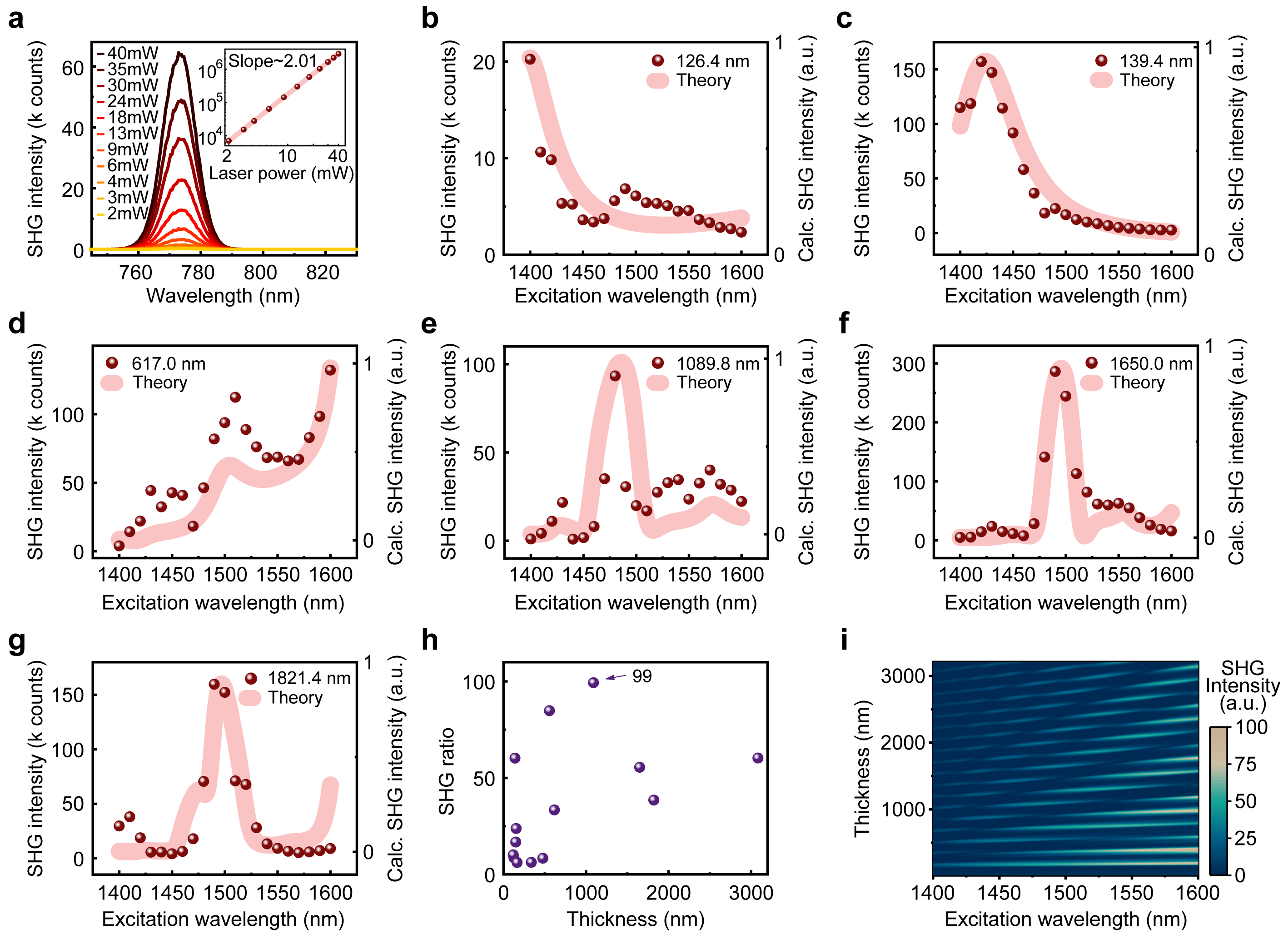} 
    \caption{
    \textbf{Giant cavity-induced modulation and complex spectral reshaping of SHG.} 
    \textbf{a} SHG spectra recorded at different excitation powers under a fundamental wavelength of 1550~nm. The inset presents the power-dependent SHG intensity on a double-logarithmic scale, giving a fitted slope of 2.01 $\pm$ 0.01. 
    \textbf{b--g} Experimental (scatter points) versus theoretical (solid lines) SHG excitation spectra for representative flakes. Their calibrated thicknesses are 126.4~nm (\textbf{b}), 139.4~nm (\textbf{c}), 617.0~nm (\textbf{d}), 1089.8~nm (\textbf{e}), 1650.0~nm (\textbf{f}), and 1821.4~nm (\textbf{g}). Measurements were performed from 1400 to 1600~nm in 10-nm steps.
    \textbf{h} SHG intensity ratio (maximum divided by minimum intensity across the measured excitation wavelengths) as a function of sample thickness. A constant incident excitation power was maintained for all recorded data in panels (\textbf{b--h}). 
    \textbf{i} Theoretical 2D contour map of the calculated SHG intensity as a function of excitation wavelength (1400 to 1600~nm) and flake thickness (up to 3210~nm), illustrating the complex topology of intertwined resonance branches.}
    \label{fig:figure3}
\end{figure*}

Having established a robust and precise thickness calibration, we shifted our focus to the nonlinear optical responses within the 3R-MoS$_2$ microcavity. 
We utilized a tunable femtosecond pulsed laser to drive these higher-order processes (see Methods for detailed optical setups).
As an initial verification of the emission, we examined the excitation power dependence.
On a double-logarithmic scale, the measured signal intensity exhibits a linear power-law scaling with a slope of 2.01 $\pm$ 0.01 (inset of Fig.~\ref{fig:figure3}a). 
This quadratic dependence confirms that the collected emission originates from a pure SHG process.

We then systematically mapped the experimental SHG excitation spectra across a broad wavelength range for several representative calibrated thicknesses (Fig.~\ref{fig:figure3}b--g; see Supplementary Information Sec.~4 for more results).
The measured spectra reveal that the SHG resonance peaks continuously shift and scale with varying cavity length.
For relatively thin flakes (e.g., 126.4 and 139.4~nm), the SHG intensity exhibits a straightforward, broadly modulated dependence on the excitation wavelength.
However, as the sample thickness increases into the micrometer regime, the wavelength-dependent evolution becomes significantly more complex, as depicted in Fig.~\ref{fig:figure3}d--g. 
In these thicker microcavities, additional resonance peaks clearly emerge alongside the primary fundamental-resonance-induced peaks.

To intuitively quantify the substantial modulation depth induced by the microcavity across these samples, we define the SHG intensity ratio for each sample as $\text{Ratio} = \frac{\max \left[ I_{\text{SHG}}(\lambda) \right]}{\min \left[ I_{\text{SHG}}(\lambda) \right]}$, where $I_{\text{SHG}}(\lambda)$ is the measured intensity at the excitation wavelength $\lambda$.
As shown in Fig.~\ref{fig:figure3}h, the experimental results reveal a stark contrast across different thicknesses, achieving an SHG intensity ratio of up to 99.
This maximum modulation is observed for a 1089.8-nm-thick flake, driven by a pronounced resonance peak at 1480~nm and a deep minimum at 1440~nm.
This giant ratio provides compelling evidence: the 3R-MoS$_2$ FP microcavity imposes  pronounced frequency selectivity and a giant modulation depth on the nonlinear optical response.

To understand the origin of this profound spectral reshaping and giant enhancement, we examine the interaction of these cavity modes.
The primary peaks are driven by the resonant enhancement of the fundamental wave, which sharply amplifies the localized electric field.
Concurrently, the generated harmonic field undergoes its own cavity resonance, producing additional spectral features.
As depicted in our theoretical 2D evolution map (Fig.~\ref{fig:figure3}i), the continuous interplay between the fundamental and second-harmonic FP resonance manifests as a complex topology of intertwined resonance branches.

A direct comparison between these theoretical calculations and our experimental spectra yields excellent overall agreement (Fig.~\ref{fig:figure3}b--g).
The model precisely captures the complex mode evolution and the specific emergence of both the primary fundamental-wave-driven peaks and the additional peaks from the second-harmonic FP resonance.
While the resonance peak positions align highly accurately, we note minor deviations in the absolute intensity envelopes.
We attribute these discrepancies primarily to the intrinsic dispersion of 3R-MoS$_2$'s second-order nonlinear susceptibility ($\chi^{(2)}$).
This wavelength-dependent property introduces variations in conversion efficiency that are not exhaustively parameterized in the current model.
Nevertheless, the precise replication of the complex oscillatory features unambiguously confirms the model's reliability in decoupling cavity interference effects from intrinsic nonlinear material properties.

\subsection{Absorption-limited Spectral Responses of THG and Short-Wavelength SHG.}

\begin{figure*}[htbp]
    \centering
    \includegraphics[width=\textwidth]{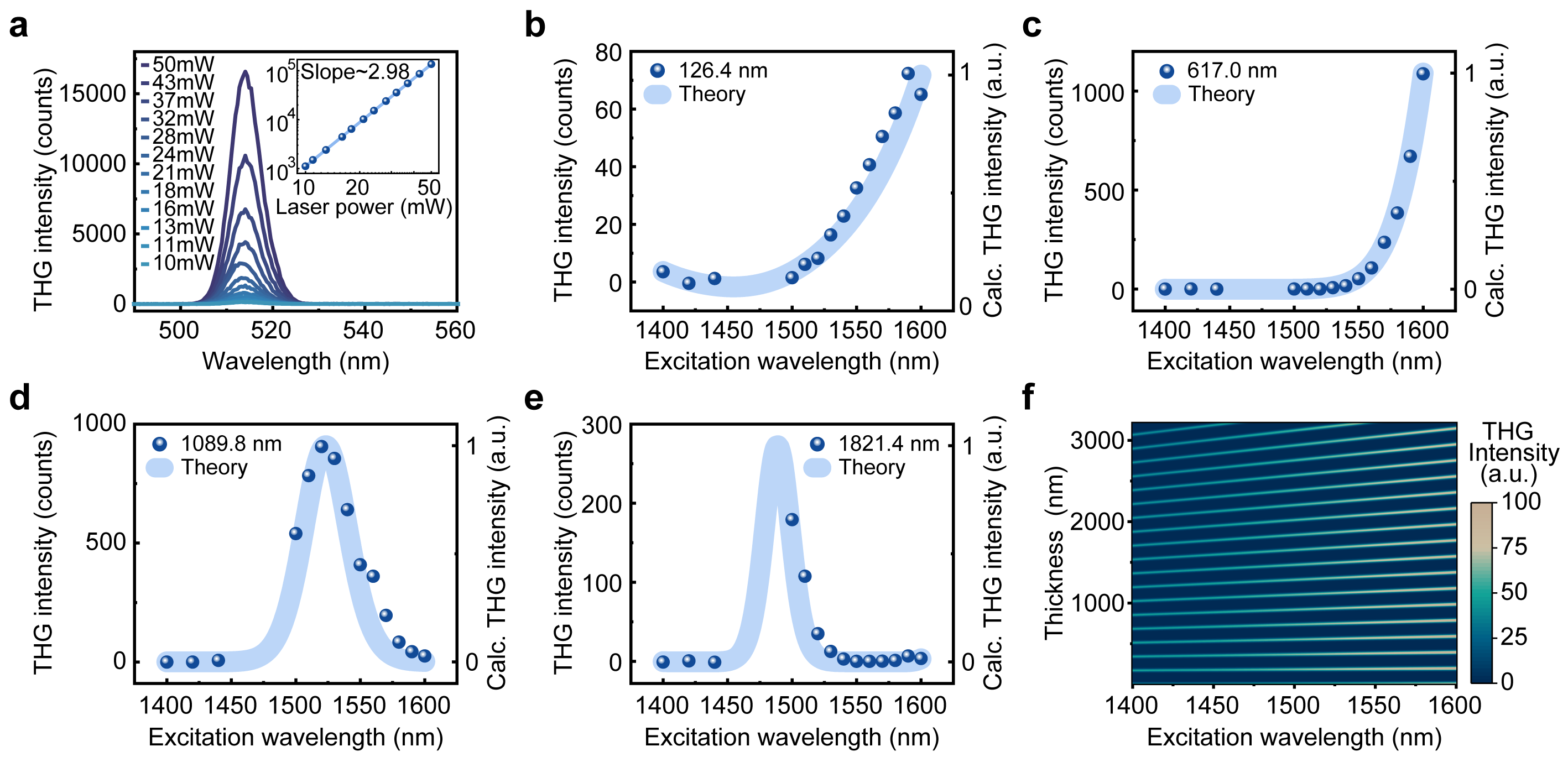} 
    \caption{
    \textbf{Absorption-limited cavity modulation of THG.} 
    \textbf{a} THG spectra recorded at different excitation powers under a fundamental wavelength of 1550~nm. The inset presents the power-dependent THG intensity on a double-logarithmic scale, yielding a fitted slope of 2.98 $\pm$ 0.01. 
    \textbf{b--e} Comparison of measured THG excitation spectra (scatter points) and theoretical predictions (solid lines) for the four representative flakes from the preceding SHG study.
    Their calibrated thicknesses are 126.4~nm (\textbf{b}), 617.0~nm (\textbf{c}), 1089.8~nm (\textbf{d}), and 1821.4~nm (\textbf{e}). 
    Left and right axes denote experimental and theoretical values, respectively.
    Measurements were performed at 1400, 1420, and 1440~nm, and across 1500--1600~nm (10-nm steps) using a constant excitation power. 
    The 1450--1490~nm range is excluded as the laser's maximum output in this band falls below the required excitation power target.
    \textbf{f} 2D contour map of the theoretically calculated THG intensity as a function of excitation wavelength and continuous flake thickness, illustrating the absorption-limited modulation governed almost entirely by the fundamental resonance.}
    \label{fig:figure4}
\end{figure*}

Having fully revealed the reshaping of the second-order nonlinear process by FP interference, we further explored the modulation capability of this microcavity on the higher-order THG response ($3\omega$).
Because THG upconverts the fundamental wave to a significantly higher photon energy, the generated photons are subjected to entirely different material dispersion and absorption regimes compared to SHG. 
This profound distinction offers a powerful probe for cavity-modulated nonlinear responses.
To rigorously confirm the physical origin of the emission, we first examined its excitation power dependence (Fig.~\ref{fig:figure4}a).
On a double-logarithmic scale, the measured signal intensity exhibits a strict cubic scaling with a fitted slope of 2.98 $\pm$ 0.01. 
This conforms precisely to theoretical expectations, confirming that the detected signals originate from a pure third-order nonlinear optical process.

Subsequently, we systematically recorded the experimental THG spectra. 
To compare the microcavity's modulation across different nonlinear orders, we deliberately present the exact same four representative samples used in the preceding SHG study.
These flakes have calibrated thicknesses of 126.4, 617.0, 1089.8, and 1821.4~nm (Fig.~\ref{fig:figure4}b--e; corresponding to Fig.~\ref{fig:figure3}b, d, e, and g, respectively; for more details, see Supplementary Information Sec.~4).
The spectral gap between 1450 and 1490~nm arises from a localized power drop in our tunable pulse laser system, which provided insufficient pump intensity to drive the third-order process.
Intriguingly, unlike the complex, multi-peak evolution observed in SHG, the THG spectra exhibit a remarkably straightforward trend, characterized by a single, well-defined resonance peak within the measured excitation windows.

To understand these spectral features, we extended our theoretical interference model to the THG regime, mapping the evolution of THG intensity as a function of excitation wavelength and sample thickness (Fig.~\ref{fig:figure4}f).
A direct comparison between the theoretically predicted curves (solid lines) and our experimentally measured THG spectra (scatter points) in Fig.~\ref{fig:figure4}b--e demonstrates excellent agreement.
Furthermore, a direct comparison with the SHG contour map (Fig.~\ref{fig:figure3}i) reveals an overall similarity in their oscillation periodicities.
This shared periodicity indicates that the dominant driving force for both processes is the resonant enhancement of the localized fundamental pump field by the geometric FP cavity modes.
However, despite this common driving factor, the THG map notably lacks the additional resonance branches observed in the SHG spectra.

The physical origin of this differential modulation behavior lies in the strongly wavelength-dependent absorption intrinsic to the 3R-MoS$_2$ crystal. 
Within our fundamental excitation range (1400 to 1600~nm), the generated SHG photons (700 to 800~nm) lie below the optical bandgap of 3R-MoS$_2$, experiencing relatively weak material absorption.
As previously established, this weak dissipation allows the sub-bandgap SHG response to exhibit a complex modulation driven by the interplay of both fundamental and harmonic FP modes.
In stark contrast, the generated THG photons (466 to 533~nm) possess energies well above the material's optical bandgap, subjecting them to severe intrinsic absorption. 
Such dissipation heavily damps the third-harmonic FP modes.
Consequently, the THG process is reduced to an absorption-limited regime, where the nonlinear signal is modulated almost exclusively by the localized enhancement of the fundamental FP resonance.

Based on this absorption-dominated mechanism, we can formulate a predictive physical picture: if the microcavity is excited at a shorter fundamental wavelength, such as 1030~nm, the generated SHG signal ($\sim$ 515~nm) is pushed above the optical bandgap and into the regime of severe intrinsic absorption.
Consequently, we predict that the spectral modulation of this high-energy SHG should closely parallel the absorption-limited THG behavior.
Dictated by severe dissipation, the emission should exhibit a streamlined modulation profile characterized by prominent enhancement exclusively where the fundamental wave satisfies the FP resonance condition.
This prediction is confirmed by our additional experimental measurements (Supplementary Information Sec.~5).
Ultimately, these findings establish that nonlinear harmonic generation in van der Waals microcavities transcends simple geometric confinement. 
Rather, it represents a highly tunable response governed by a delicate interplay among the intrinsic material absorption, the FP effects at the fundamental frequency, as well as those at the harmonic frequencies.
This interplay offers a powerful new lever for engineering wavelength-selective, on-chip nanophotonic devices.

\section*{Discussion}

In summary, we have systematically investigated the profound influence of FP modes on both SHG and THG within wavelength-scale 3R-MoS$_2$ microcavities.
By establishing a highly precise geometric baseline via non-destructive broadband reflectance, we created a self-consistent linear-to-nonlinear experimental framework.
Building on this rigorous in situ metrology, our results unambiguously demonstrate that these internal resonances induce a massive modulation depth, achieving an SHG enhancement of over two orders of magnitude. 
Furthermore, our cross-harmonic comparative analysis reveals that the microcavity's nonlinear emission is consistently governed by three distinct factors: the intrinsic material absorption, as well as fundamental and harmonic FP effects.
We demonstrate that the spectral evolution of these processes diverges inherently depending on the material's intrinsic dissipation.
Sub-bandgap emission is shaped by the complex interplay of both fundamental and harmonic FP modes, while above-bandgap emission is reduced to an absorption-limited regime.
Under severe intrinsic dissipation, the FP resonance of the harmonic field is strongly suppressed, dictating that the nonlinear emission is modulated predominantly by the fundamental resonances.

Looking forward, these findings provide a robust theoretical and experimental framework for the development of next-generation van der Waals nanophotonics. 
We have demonstrated the ability to decouple, independently modulate, and tailor distinct nonlinear optical signals.
Crucially, this is achieved within a simple, unpatterned monolithic two-dimensional microcavity.
This capability opens exciting new avenues for designing highly compact, multi-band nonlinear light sources and advanced on-chip optical devices.

\section*{Methods}
\noindent\textbf{Sample fabrication} 

\noindent The 3R-MoS$_2$ flakes were mechanically exfoliated from bulk crystals (HQ Graphene) onto a fused silica substrate that was first cleaned using a vacuum plasma cleaner (SUNJUNE PLASMA VPR3). 
The thicknesses of the exfoliated flakes were characterized by AFM (Bruker nanoIR3s). \\

\noindent\textbf{Optical measurements}

\noindent Linear optical characterization (broadband reflectance spectroscopy) was performed using a custom-built confocal microscope (modified from Fig.~S6) equipped with a halogen lamp as the light source. The visible range (600--965~nm) was collected using a 20$\times$ Nikon objective (NA = 0.45) and recorded via a grating spectrometer (HRS-300SS-NI, Teledyne Princeton Instruments) coupled with a PyLoN CCD camera. For the infrared range (985--1570~nm), a 40$\times$ reflective objective (LMM40X-P01, Thorlabs, NA = 0.5) was employed, with detection by a PyLoN-IR CCD camera.

\noindent Nonlinear optical measurements (SHG and THG) were carried out using a custom-built system (Fig.~S6) with a femtosecond laser (Levante IR fs, APE GmbH) as the excitation source. The laser beam was focused onto the sample surface via a 5$\times$ Mitutoyo objective (NA = 0.14). Both SHG and THG signals were collected by the same objective, isolated through bandpass filters, and detected by the aforementioned spectrometer coupled with the PyLoN CCD camera. All measurements were conducted under ambient conditions.\\

\noindent\textbf{Numerical simulations} 

\noindent To quantitatively model the complex optical responses within the 3R-MoS$_2$ microcavity, our numerical framework was divided into linear and nonlinear regimes.
The linear broadband reflectance spectra of the air/3R-MoS$_2$/SiO$_2$ substrate system were calculated utilizing the TMM \cite{rumpf2011improved}. 
This approach rigorously accounted for the multiple interfacial reflections and the phase accumulation across the flake's geometric thickness.

\noindent Building upon the fundamental electric field distributions ($E^\omega$) established by the linear cavity modes, the nonlinear optical responses were calculated using a rigorous electromagnetic Green's function method.
Because our experimental configuration collected the backward-propagating harmonic signals (i.e., the nonlinear reflection spectra), the observation point for the Green's function ($G$) was positioned at the upper air--3R-MoS$_2$ interface. 
The generated nonlinear optical intensity, $I^{\text{NL}}$, was directly derived from the coherent spatial integration of the nonlinear polarization $P^{\text{NL}}$ convoluted with the Green's function across the thickness of the film system:
\begin{equation}
    I^{\text{NL}} \propto \left| E^{\text{NL}} \right|^2 \propto \left| \int G \cdot P^{\text{NL}} \, dz \right|^2
\end{equation}
\noindent where $E^{\text{NL}}$ represents the reflected nonlinear electric field. 
The nonlinear polarization source terms driving the SHG and THG processes were governed by the fundamental cavity field $E^{\omega}$, and were expressed respectively as:
\begin{equation}
    P^{2\omega} = \chi^{(2)} (E^\omega)^2
\end{equation}
\begin{equation}
    P^{3\omega} = \chi^{(3)} (E^\omega)^3
\end{equation}
\noindent Here, $\chi^{(2)}$ and $\chi^{(3)}$ denote the effective second- and third-order nonlinear susceptibilities of the 3R-MoS$_2$ material. 
By coupling the TMM-derived internal field $E^\omega$ with this Green's function formalism, we accurately simulated how the fundamental and second-harmonic FP resonance and intrinsic absorption collectively dictated the final harmonic generation efficiency.
The wavelength-dependent dielectric constants used in these simulations were referenced from established literature for both the MoS$_2$ \cite{MoS2_permittivity} and the SiO$_2$ substrate \cite{malitson1965interspecimen}.

\section*{Data availability}
The data supporting this study are available upon reasonable request.

\section*{Acknowledgements} 
The authors acknowledge support from the National Key R{\&}D Program of China (2024YFB2808100), the National Natural Science Foundation of China (62475194, 62305249, 12574364, 62192770, 62192772),  the Science and Technology Commission of Shanghai Municipality (23190712300, 23ZR1465800), the National Research Foundation of Korea (NRF) funded by the Korean government (MSIP) (RS-2020-NR049536, RS-2025-25443562). 

\section*{Author Contributions}
R. S., Z. C., and J. X. contributed equally to this work. T. J. conceived and designed the experiments. R. S. fabricated the samples and conducted the far-field measurements with the help of Z. R. W., Z. T. W., S. Z., W. S., and Z. P.. J. X. performed the calculations and simulations. R. S., Z. C., and J. X. contributed to the data analysis and manuscript writing with the help of all authors.
T. J. and D. H. supervised the entire project. All authors discussed and interpreted the results.

\section*{Competing interests}
\noindent The authors declare no competing interests.

\section*{Supplementary Information}
\textbf{This PDF file includes:} 

Sections S1 to S6

Figs. S1 to S6

\bigskip

\bibliographystyle{naturemag}
\bibliography{main-ref}
\end{document}